\documentclass[superscriptaddress,reprint, prl]{revtex4-1}
\usepackage{graphicx}
\usepackage{amsmath}
\usepackage{amssymb}
\usepackage[squaren, thinqspace]{SIunits}
\usepackage{xspace}
\usepackage{setspace}
\usepackage{color}
\usepackage{units}
\usepackage{textcomp}
\usepackage{amssymb}
\usepackage{textcomp}
\usepackage{pifont}

\begin{document}
\title{Experimental observation of an enhanced anisotropic magnetoresistance in non-local configuration}
%\date{2011-MAY-25}

\author{D.\,R\"{u}ffer}\thanks{now at Laboratoire des Mat\'{e}riaux Semiconducteurs,  Institut des Mat\'{e}riaux, Ecole Polytechnique F\'{e}d\'{e}rale de Lausanne,  CH-1015 Lausanne,  Switzerland}
\affiliation{Walther-Mei{\ss}ner-Institut, Bayerische Akademie der Wissenschaften, D-85748 Garching, Germany}

\author{F.\,D.\,Czeschka}
\affiliation{Walther-Mei{\ss}ner-Institut, Bayerische Akademie der Wissenschaften, D-85748 Garching, Germany}

\author{R.\,Gross}

\affiliation{Walther-Mei{\ss}ner-Institut, Bayerische Akademie der Wissenschaften, D-85748 Garching, Germany}
\affiliation{Physik-Department, Technische Universit\"{a}t M\"{u}nchen, D-85748 Garching, Germany}

\author{S.\,T.\,B.\,Goennenwein}

\email{goennenwein@wmi.badw.de}
\affiliation{Walther-Mei{\ss}ner-Institut, Bayerische Akademie der Wissenschaften, D-85748 Garching, Germany}

\begin{abstract}

We compare non-local magnetoresistance measurements in multi-terminal Ni nanostructures with corresponding local experiments. In both configurations, the measured voltages show the characteristic features of anisotropic magnetoresistance (AMR). However, the magnitude of the non-local AMR signal is up to one order of magnitude larger than its local counterpart. Moreover, the non-local AMR increases with increasing degree of non-locality, i.e., with the separation between the region of the main current flow and the voltage measurement region. All experimental observations can be consistently modeled in terms of current spreading in a non-isotropic conductor. Our results show that current spreading can significantly enhance the magnetoresistance signal in non-local experiments. 

\end{abstract}
\maketitle

Non-local voltage measurements are an important tool in solid state physics, e.g., for the study of spin accumulation and spin currents in nanoscale ferromagnet/normal metal hybrid samples \cite{non-local-spin-injection:JohnsonSilsbee:PRL:1985,Jedema:2001,seki_giant_2008,mihajlovic_negative_2009,spin-pumping:YIG:Kajiwara:Nature,kimura_room-temperature_2007}.
The term 'non-local' hereby means that the region of the sample probed by the voltage measurement nominally is free of charge current. In other words, charge current related effects can be suppressed in appropriate non-local voltage measurements \cite{Tang:in:Awschalom-Halbleiter-Spin-Buch}.
In contrast, in a conventional `local' measurement, the voltage is probed along the region of charge current flow, such that galvanic effects due to charge motion usually are dominant.
However, in spite of the widespread use of non-local measurements for the study of ferromagnetic/non-magnetic hybrid devices, little is known about non-local effects in plain ferromagnetic nanostructures.
We here investigate the anisotropic magnetoresistance (AMR) in ferromagnetic $3d$ transition metal nanostructures, comparing measurements in both local and non-local geometry. Both the local and the non-local signals show the $\cos^2\theta$ dependence characteristic for AMR~\cite{McGuire:MTr-AMR:IEEETM:1975}, where $\theta$ is the angle enclosed by the magnetization of the ferromagnet and the direction of the local current flow. 
However, while non-local signals usually decay exponentially with the separation $L$  between the local current flow and the voltage probes, we find that in our case, the magnitude of the non-local AMR (the relative resistance change) increases linearly with increasing separation $L$. Consequently, the non-local AMR is up to one order of magnitude larger than its local counterpart, measured in one and the same device. 
We show that current spreading in combination with a non-isotropic resistivity tensor indeed accounts for the substantial increase of the non-local AMR magnitude with increasing degree of non-locality.

\begin{figure}
\begin{centering}
\includegraphics[width=8cm]{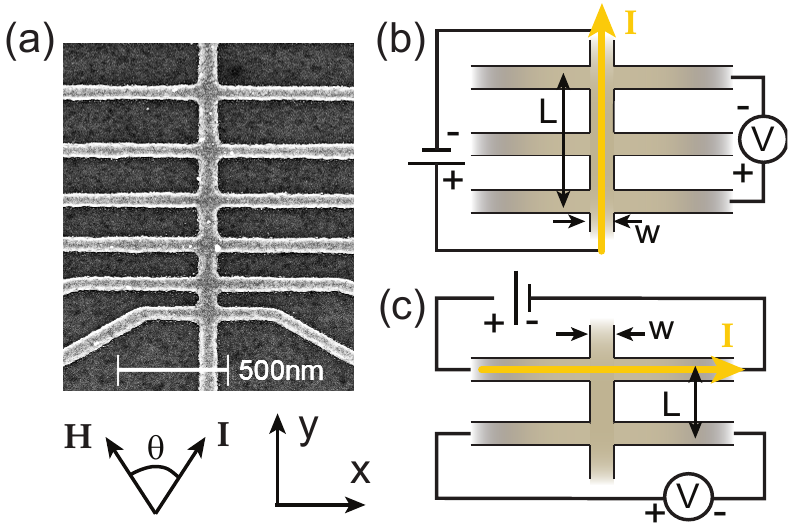}
\par\end{centering}
\caption{(a) SEM picture of a typical Hall bar sample. (b),(c) Contact schematic for (b) local and (c) non-local measurement configurations.
\label{fig:Geometry}}
\end{figure}

The ferromagnetic metal Hall bar structures were deposited onto SiO$_x$/Si wafers via electron-beam lithography, electron-beam evaporation and lift-off. Fig.~\ref{fig:Geometry}(a) shows a typical SEM image of a typical sample.
We have fabricated and studied samples made from Ni with different film thicknesses, and found that the film thickness has no significant impact on the results. For the sake of simplicity, we here focus on one particular sample with a Ni film thickness of $\unit[50]{nm}$, a main Hall bar width $w=\unit[110]{nm}$ (vertical wire in Fig.~\ref{fig:Geometry}(a)), and $\unit[80]{nm}$ wide voltage leads (horizontal wires). The distance $L$ between adjacent voltage leads, measured from center to center, systematically varies from $\unit[150]{nm}$ to $\unit[300]{nm}$.

For the resistance measurements, an alternating (switched) DC current of $I=\unit[\pm100]{\mu A}$ was applied, and the resulting DC voltage was recorded with a nanovoltmeter. For such current magnitudes, the $I$-$V$ characteristic of the Hall bar device is linear. More precisely, the measured resistivity changes by less than $\unit[1]{\%}$ (e.g. by Joule heating) for measurement current magnitudes $\unit[10]{\mu A}\le I \le \unit[500]{\mu A}$. The local quantities, i.e., the longitudinal resistances $R$ and the corresponding resistivity $\rho$, are determined using a conventional four point measurement configuration (Fig.~\ref{fig:Geometry}(b)). The current is sent through the Hall bar, while the voltage probes are attached to two horizontal voltage leads. The high and low voltage and current connections, symbolized by $+$ and $-$ in Fig.~\ref{fig:Geometry}(b), respectively. 
Figure~\ref{fig:Geometry}(c) illustrates  the configuration used for the detection of the \emph{non-local} voltage signals. The current is passed through one of the horizontal wires. The voltage probes are situated on the ends of another, parallel wire, which is not directly exposed to the current flow. We find that the non-local voltages $V_{\mathrm{NL}}$ scale linearly with the applied local current. We therefore use the \emph{non-local resistance} $R_{\mathrm{NL}}=V_{\mathrm{NL}}/I$ in the following. One nevertheless should keep in mind that $R_{\mathrm{NL}}$ is not a true resistance but a normalized voltage signal.

The magnetoresistance measurements were conducted with an external magnetic field $\mathbf{H}$ applied in the substrate plane. We here use the angle $\theta$ enclosed between $\mathbf{H}$ and the current direction $\mathbf{I}$ to quantify the field orientation. Note that the direction of current flow is different in the local and the non-local geometry (cf.~Fig.~\ref{fig:Geometry}). For local voltage measurements, $\theta=\unit[0]{\text{\textdegree}}$ corresponds to $\mathbf{H}\parallel\mathbf{y}$, since $\mathbf{I}\parallel\mathbf{y}$ in this case. In contrast, $\theta=\unit[0]{\text{\textdegree}}$ refers to $\mathbf{H}\parallel\mathbf{x}$ for non-local experiments. All experiments discussed in the following were performed as a function of $\theta$ at $\mu_0 |\mathbf{H}| = \unit[2]{T}$. Since this magnetic field magnitude is several times larger than the Ni anisotropy fields, the magnetization $\mathbf{M}$ of the Ni nanostructure will always be aligned along $\mathbf{H}$ in good approximation ($\mathbf{M}||\mathbf{H}$). We therefore employ $\theta$ also for the discussion of the anisotropic magnetoresistance (AMR) measurements below, although the AMR strictly speaking is governed by the relative orientation of $\mathbf{M}$ with respect to $\mathbf{I}$. In particular, we will use $R_{||}$ for the resistance measured at $\theta=\unit[0]{\text{\textdegree}}$ ($\mathbf{M}||\mathbf{I}$), and $R_{\perp}$ for $\theta=\unit[90]{\text{\textdegree}}$ ($\mathbf{M}\perp\mathbf{I}$). All measurements presented in the following were taken at room temperature. We also studied the local and non-local AMR at $T=\unit[3]{K}$, and observed AMR magnitudes comparable to $T=\unit[300]{K}$.

\begin{figure}
\begin{centering}
\includegraphics[width=8cm]{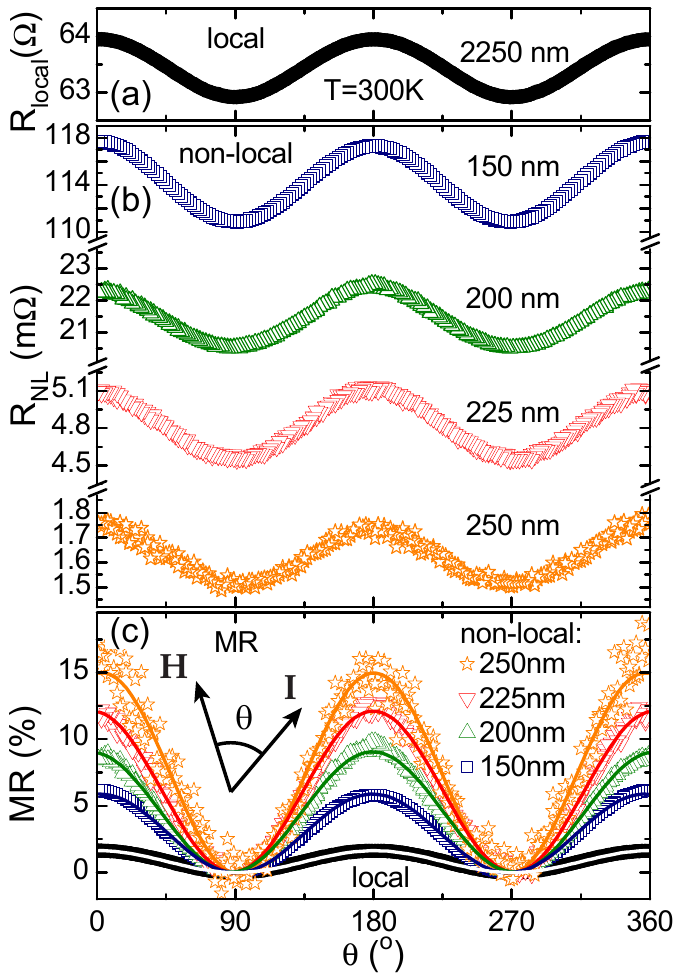}
\par\end{centering}
\caption{(a) Local and (b) non-local resistance of a $\unit[50]{nm}$ thick Ni Hall bar recorded at room temperature as a function of the orientation $\theta$ of an externally applied magnetic field $\mu_{0}H=\unit[2]{T}$ for (a) $L=\unit[2250]{nm}$ and (b) $L$ ranging between 150 and $\unit[250]{nm}$ (cf.~Fig.~\ref{fig:Geometry}). (c) The $\mathrm{MR}\left(\theta\right)= (R(\theta)-R(\theta=\unit[90]{\text{\textdegree}}))/R(\theta=\unit[90]{\text{\textdegree}})$ increases by up to one order of magnitude with the degree of non-locality, i.e., with the separation $L$ in the non-local experiments. The full lines are fits to the data according to Eq.~(\ref{eq-AMR-NONlocal}).
\label{fig:AMR-measurement}}
\end{figure}

The resistivity $\rho_{\parallel}$ of the Ni wires was determined to $\unit[\unit{17.0}]{\mbox{\textmu}\Omega\, cm}$ at  $T=\unit[300]{K}$. Due to the AMR effect,
$R(\theta)=R_{\perp}+\left(R_{||}-R_{\perp}\right)\cos^{2}\theta$ characteristically changes as a function of $\theta$~\cite{McGuire:MTr-AMR:IEEETM:1975}, see Fig.~\ref{fig:AMR-measurement}(a). As expected for Ni~\cite{campbell_spontaneous_1970,Ohandley:ModernMagneticMaterials}, $R_{||}>R_{\perp}$. The maximum relative resistance change, the so-called AMR value, is commonly defined as $\mathrm{AMR}=\left(R_{||}-R_{\perp}\right)/R_{\perp}$. We find an $\mathrm{AMR}=\unit[1.64]{\%}$ at $\unit[300]{K}$ in the present sample, which is comparable to literature values for Ni \citep{jia_effect_1997}.
AMR measurements in the non-local geometry are shown in Figs.~\ref{fig:AMR-measurement}(b) for different  $L$. They all obey
\begin{equation}\label{eq-AMR-NONlocal}
 R_{\mathrm{NL}}(\theta)=R_{\mathrm{NL,\perp}}+\left(R_{\mathrm{NL},||}-R_{\mathrm{NL},\perp}\right)\cos^{2}\theta
\end{equation}
as expected for an AMR signal, and are maximal for parallel alignment of current and magnetic field and minimal for a perpendicular orientation. In other words, the qualitative behavior of $R_{\mathrm{NL}}(\theta)$ is similar to the local resistance dependence $R(\theta)$. Note also that $\unit[1.5]{m\Omega}\leq R_{\mathrm{NL}}\leq\unit[1.75]{m\Omega}$ at $L=\unit[250]{nm}$  is nearly two orders of magnitude smaller than $\unit[111]{m\Omega}\leq R_{\mathrm{NL}}\leq\unit[117]{m\Omega}$ at $\unit[150]{nm}$. This corroborates the notion that $R_{\mathrm{NL}}$ decays exponentially with the separation $L$ from the current.
Interestingly, however, the magnitude of the non-local AMR value $\mathrm{AMR}_{\mathrm{NL}}=\left(R_{\mathrm{NL},||}-R_{\mathrm{NL},\perp}\right)/R_{\mathrm{NL},\perp}$ \emph{increases} with increasing $L$. This is directly evident from Fig.~\ref{fig:AMR-measurement}(c), in which the magnetoresistance $\mathrm{MR}\left(\theta\right)= (R(\theta)-R_{\perp})/R_{\perp}$ is plotted both for the local resistance $R(\theta)$ (full symbols), and for several non-local configurations with different $L$ (colored open symbols). Fits using Eq.~(\ref{eq-AMR-NONlocal}) are represented by the solid lines. Figure \ref{fig:AMR-measurement}(c) in particular also shows that the AMR in the non-local geometry exceeds its local counterpart, by up to one order of magnitude for large $L$.

\begin{figure}
\begin{centering}
\includegraphics[width=8cm]{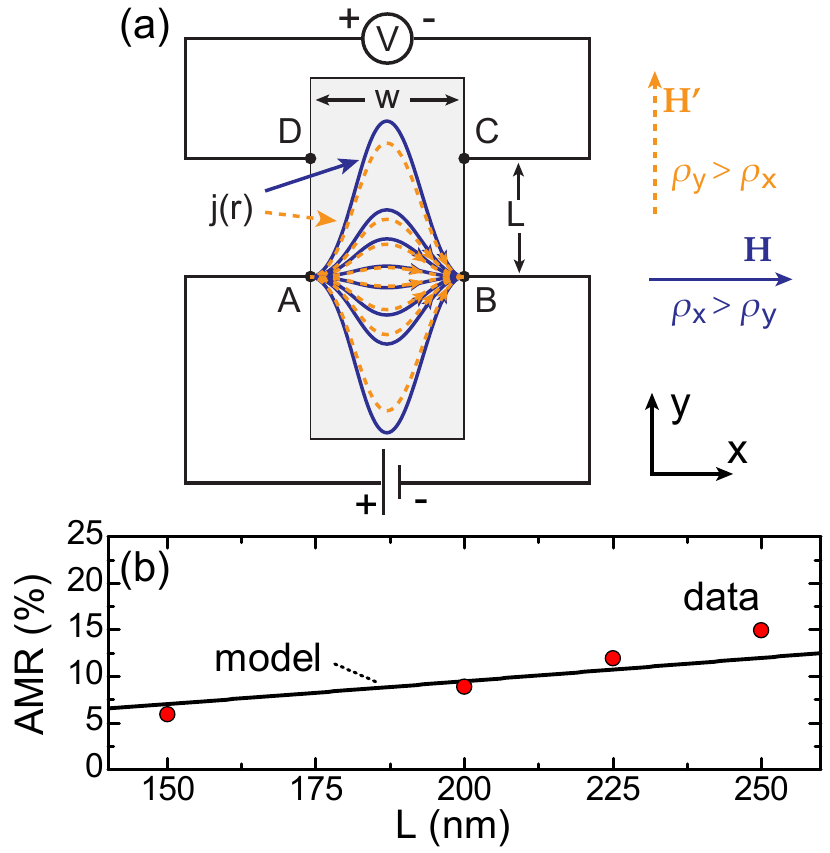}
\par\end{centering}
\caption{(a) Sketch of current spreading in the presence of anisotropic magnetoresistance. For the magnetic field parallel to the main current direction and thus parallel to the $\mathbf{x}$-axis (blue solid lines), the resistivity is lower in $\mathbf{y}$-direction. This leads to an enhanced current spreading along $\mathbf{y}$. In contrast, current spreading along $\mathbf{y}$ is decreased for $\mathbf{H}\perp\mathbf{I}$ (orange dashed lines). (b) The non-local AMR increases with the separation $L$ between the main current path and the voltage detection wire. The symbols represent the experimental $\mathrm{AMR}_{\mathrm{NL}}$ values for $\unit[300]{K}$. The solid line depicts the $\mathrm{AMR}_{\mathrm{NL}}(L)$ according to the analytical model (Eq.~(\ref{eq:ni:AMR-L})).}
\label{fig:simu_L-dependence}
\end{figure}

The increase of $\mathrm{AMR}_{\mathrm{NL}}$ with $L$ observed in experiment can be quantitatively understood in terms of current spreading in a non-isotropic conductor. Due to the AMR effect, the Ni film plane contains both low and high resistance directions. Moreover, these directions change as a function of the orientation of the magnetic field, such that also the current paths will spread differently for different $\theta$ (cf.~Fig.~\ref{fig:simu_L-dependence}(a)). In Ni, the resistance is lower for current perpendicular to the magnetic field. Thus, for field in $x$-direction (blue solid lines in Fig.~\ref{fig:simu_L-dependence}(a)), the current can spread ``easier'' along $y$, which results in more current and thus also a higher voltage difference in the region of the non-local voltage probes, as compared to the situation for field parallel to the $y$-direction (orange dashed lines), where the $y$-direction is high resistance. Since the voltage decays exponentially, the difference between the two field configurations is getting more and more pronounced with increasing $L$, leading to an enhanced $\mathrm{AMR}_{\mathrm{NL}}$ value.

The current spreading can be modeled quantitatively using a two dimensional resistivity tensor $\hat{\rho}$ depending on $\theta$,
\begin{equation}
\hat{\rho}=\left(\begin{array}{cc}
\rho_{\perp}+\Delta\rho\cos^{2}\theta & \Delta\rho\sin\theta\cos\theta\\
\Delta\rho\sin\theta\cos\theta & \rho_{\parallel}-\Delta\rho\cos^{2}\theta\end{array}\right),
\label{eq:resistivity}
\end{equation}
with $\Delta\rho=\rho_{\parallel}-\rho_{\text{\ensuremath{\perp}}}$. Considering a sample with width $w$, thickness $t$ and point like current injection probes, which are remote ($L\gg w$) from the voltage probes, one can calculate the expected non-local resistance using the van-der-Pauw theorem,  generalized to anisotropic media~\cite{price_extension_1972,kleiza_extension_2007}:
\begin{equation}
\exp\left(-\pi tsR_{\mathrm{AB,CD}}\right)+\exp\left(-\pi tsR_{\mathrm{BC,DA}}\right)=1.
\label{eq:gen-vdP}
\end{equation}
Here, $s=\sqrt{\det\hat{\sigma}}$, and $\hat{\sigma}=\hat{\rho}^{\mbox{-}1}$ is the conductivity tensor. The resistance $R_{\mathrm{AB,CD}}$ is defined as the voltage difference $V_{\mathrm{D}}-V_{\mathrm{C}}$ between points D and C per current $I_{A\rightarrow B}$ from contacts A to B, and $R_{\mathrm{BC,DA}}$ denotes the respective permutation. Using the notation of  Fig.~\ref{fig:simu_L-dependence}(a), $R_{\mathrm{AB,CD}}$ corresponds to the non-local resistance $R_{\mathrm{NL}}$. For $L\gg w$, the local longitudinal resistance $R=\rho L/(wt)$ will be equal to  $R_{\mathrm{BC,DA}}=\left(V_{\mathrm{A}}-V_{\mathrm{D}}\right)/I_{\mathrm{B\rightarrow C}}$. Considering that $s=\sqrt{\rho_{\parallel}\rho_{\perp}}$ for the resistivity tensor of Eq.~\eqref{eq:resistivity} and introducing  $\gamma=\rho_{\parallel}/\rho_{\perp}=1+\mathrm{AMR}$,  Eq.~\eqref{eq:gen-vdP} reads
\begin{equation}
R_{\mathrm{NL}}\left(L,\theta\right)=\frac{s}{\pi t}\exp\left(-\frac{\pi  L \sqrt{\gamma}}{w}\left[1-\left(1-\gamma^{-1}\right)\cos^{2}\theta\right]\right).
\label{eq: RNL_L_theta}
\end{equation}
This yields
\begin{equation}
\mathrm{AMR}_{\mathrm{NL}}\left(L\right)=\exp\left(\frac{\pi  L}{w}\cdot\left[\gamma^{1/2}-\gamma^{-1/2}\right]\right)-1.
\label{eq:ni:AMR-L}
\end{equation}
Equation (\ref{eq:ni:AMR-L}) correctly reproduces all salient features of the experimental data. In particular, considering the geometry and the AMR value of our sample, the exponential in Eq.~(\ref{eq:ni:AMR-L}) can be expanded, yielding $\mathrm{AMR}_{\mathrm{NL}}\left(L\right)=\mathrm{AMR} \pi L /w$. This shows that the  non-local AMR magnitude indeed increases with $L$.
Figure~\ref{fig:simu_L-dependence}(b) shows $\mathrm{AMR}_{\mathrm{NL}}$ obtained from Eq.~(\ref{eq:ni:AMR-L}) using $\gamma$ as determined by local measurements. Calculation (line) and experiment (symbols) agree very well and confirm current spreading in an anisotropic conductor as the origin of our observations. Note that there are no free parameters in the calculation.

In conclusion, we have studied the local and the non-local AMR in nanoscale Ni Hall bars. Both the local and the non-local signal change with the orientation of the Ni magnetization in a fashion characteristic for AMR. However, the non-local AMR is up to one order of magnitude larger than its local counterpart, and increases roughly linearly with the degree of non-locality $L$. These experimental findings can be consistently understood in terms of current spreading in a material with anisotropic conductance. Our results thus show that magnetoresistive properties can be substantially enhanced in non-local measurements. Clearly, a more detailed investigation of non-local magneto-resistive effects due to current spreading appears desirable, in particular also in ferromagnet/normal metal hybrid structures frequently used in present experiments.

\begin{acknowledgments}
Financial support by the Deutsche Forschungsgemeinschaft via the Excellence Cluster \textquotedblleft{}Nanosystems Initiative Munich NIM\textquotedblright{} is gratefully acknowledged.
\end{acknowledgments}

%\bibliography{literature_Rueffer}{}

\begin{thebibliography}{10}

\bibitem{non-local-spin-injection:JohnsonSilsbee:PRL:1985}
M.~Johnson and R.~H. Silsbee, Phys. Rev. Lett. \textbf{55}, 1790 (1985).

\bibitem{Jedema:2001}
F.~Jedema, A.~Filip, and B.~{van Wees}, Nature \textbf{410}, 345 (2001).

\bibitem{seki_giant_2008}
T.~Seki \emph{et~al.}, Nature Materials \textbf{7}, 125 (2008).

\bibitem{mihajlovic_negative_2009}
G.~Mihajlovic \emph{et~al.}, Phys. Rev. Lett. \textbf{103}, 166601 (2009).

\bibitem{spin-pumping:YIG:Kajiwara:Nature}
Y.~Kajiwara \emph{et~al.}, Nature \textbf{464}, 262 (2010).

\bibitem{kimura_room-temperature_2007}
T.~Kimura \emph{et~al.}, Phys. Rev. Lett. \textbf{98}, 156601 (2007).

\bibitem{Tang:in:Awschalom-Halbleiter-Spin-Buch}
H.~X. Tang \emph{et~al.}, \emph{Semiconductor Spintronics and Quantum
  Computation}, edited by D.~D. Awschalom, D.~Loss, and N.~Samarth, NanoScience
  and Technology (Springer, Berlin, 2002).

\bibitem{McGuire:MTr-AMR:IEEETM:1975}
T.~R. {McGuire} and R.~I. Potter, {IEEE} Trans. Magn. \textbf{MAG-11}, 1018
  (1975).

\bibitem{campbell_spontaneous_1970}
I.~A. Campbell, A.~Fert, and O.~Jaoul, J. Phys. C. \textbf{3}, S95 (1970).

\bibitem{Ohandley:ModernMagneticMaterials}
R.~C. {O'Handley}, \emph{Modern Magnetic Materials : Principles and
  Applications} (John Wiley \& Sons, New York, 2000).

\bibitem{jia_effect_1997}
Y.~Q. Jia, S.~Y. Chou, and J.~Zhu, J. Appl. Phys. \textbf{81}, 5461 (1997).

\bibitem{price_extension_1972}
W.~L.~V. Price, J. Phys. D. \textbf{5}, 1127 (1972).

\bibitem{kleiza_extension_2007}
J.~Kleiza, M.~Sapagovas, and V.~Kleiza, Informatica \textbf{18}, 253 (2007).

\end{thebibliography}
%\bibliographystyle{PRL}
\end{document}